\documentclass[12pt]{amsart}
\usepackage{amssymb,amscd}
\usepackage{pb-diagram}
\usepackage{verbatim}
\usepackage{graphicx}

\newtheorem*{Whitney towers}{Theorem~\ref{Whitney towers}}
\newtheorem*{h-towers}{Theorems ~\ref{half} \& \ref{$(n)$-solvable}}

\newtheorem*{surgery curves}{Theorem~\ref{surgery curves}}
\newtheorem*{cg=0}{Theorem~\ref{vanish}}

\theoremstyle{definition}

\numberwithin{equation}{section}
\numberwithin{figure}{section}

\newcommand{\x}{\times}

\newcommand{\nl}{\newline}
\newcommand{\Z}{\mathbb{Z}}
\newcommand{\N}{\mathbb{N}}
\newcommand{\C}{\mathbb{C}}

\newcommand{\R}{\mathbb{R}}

\newcommand{\G}{\mathbb{G}}

\newcommand{\f}{\noindent}

\newcommand{\HH}{\mathcal{H}}

\newcommand{\DD}{\mathcal{D}}

\renewcommand{\SS}{\mathcal{S}}

\title{Supersymmetry, homology with twisted coefficients 
and $n$-dimensional knots}
\author{Eiji Ogasa}
\address{High Energy Physics Theory Group\\
Department of Physics\\
Ochanomizu University\\ 
Bunkyo-ku\\
Tokyo 112-8610\\ 
JAPAN\\}
\email{ogasa@hep-th.phys.u-tokyo.ac.jp}

\begin{document}
\begin{abstract} 
Let $n$ be any natural number. 
Let $K$ be any $n$-dimensional knot in $S^{n+2}$. 
We define a supersymmetric quantum system for $K$ 
with the following properties. 
We firstly construct a set of functional spaces 
(spaces of fermionic \{resp. bosonic\} states) 
and a set of operators (supersymmetric infinitesimal transformations) 
in an explicit way. 
Thus we obtain a set of the Witten indexes for $K$.  
Our Witten indexes are topological invariants 
for $n$-dimensional knots. 
Our Witten indexes  are not zero in general. 
If $K$ is equivalent to the trivial knot, 
all of our Witten indexes are zero.     
Our Witten indexes restrict the Alexander polynomials of $n$-knots. 
If one of our Witten indexes for an $n$-knot $K$ is nonzero, 
then one of the Alexander polynomials of $K$ is nontrivial. 
Our Witten indexes are connected with homology with twisted coefficients. 

Roughly speaking, our Witten indexes have path integral representation 
by using a usual manner of supersymmetric theory.
\end{abstract}


\maketitle
\section{Introduction and our main claims}  
\label{Introduction}  

When we discuss time-space, 
we often regard time-space as a manifold other than $\R^4$ 
especially after  Einstein's general theory of relativity \cite{Einstein}. 
This kind of discussion is essentially useful both for physics and 
for mathematics.

As is well-known, 
the papers \cite{AlvarezGaume} \cite{Witten}
state that: 
if we discuss the Witten index for a kind of supersymmetric quantum theory 
on a manifold,  
then we get the Atiyah-Singer index of an operator on some bundles 
on the manifold, 
which gives a topological invariant of the manifold in some cases.

Let $n$ be any natural number. 
Let $K$ be any $n$-dimensional knot in $S^{n+2}$ 
(see e.g. \cite{CochranOrr}\cite{LevineAlex} \cite{LevineOrr} \cite{Rolfsen} 
for $n$-knots).
In this paper we consider a question 
whether we can make a model of supersymmetric quantum theory
associated with the $n$-knot $K$ with the following properties(1)-(4): 

\begin{itemize}
\item[(1)] 
Our Witten indexes for our supersymmetric infinitesimal operators 
are topological invariants of $n$-knots $K$. 

\item[(2)] 
All of our  Witten indexes are not zero in general.

\item[(3)]  
All of our Witten indexes are zero for the trivial $n$-knot.

\item[(4)]
The space of fermionic (resp. bosonic) states in this model is 
infinite dimensional vector space.  
\end{itemize}

\f 
In \S\ref{Our} we give a positive answer. 
We construct such a model in an explicit way.   
We construct a set of functional spaces 
(spaces of fermionic \{resp. bosonic\}states), 
a set of operators (supersymmetric infinitesimal transformations, and 
Hamiltonians). Our Witten indexes have the above properties (1)-(4) and, 
furthermore, have the following properties(5)(6):  

\begin{itemize}
\item[(5)]
Our Witten indexes restrict the Alexander polynomial 
for $n$-dimensional knots.  
\item[(6)]
Our Witten indexes restrict the homology of $S^{n+2}-K$ with a twisted coefficient. 
\end{itemize}

\f The Witten indexes have  path integral representation 
(see \cite{AlvarezGaume}).
Roughly speaking, hence  
our way gives path integral representation to 
some topological invariants for $n$-knots.

The study of $n$-dimensional knots and $n$-dimensional links
has a long history and many fruitful results 
(see e.g. \cite{CochranOrr}\cite{LevineAlex} \cite{LevineOrr} \cite{Rolfsen}). 
Note that it is known that, for any natural number $n$, 
there are countably infinitely many nonequivalent $n$-knots.



\section{Review of the Witten index} \label{Review}

In order to define our Witten indexes, 
we review the Witten index 
(see \cite{AlvarezGaume}\cite{Witten}). 
But we do not mind how rigorous our explanation is 
from mathematical viewpoints.

Let $\HH$ be an infinite dimensional vector space (over $\C$)
which is a subset of a set of 
sections of a $\C^n$-vector bundle on a manifold. 
Suppose $\HH$ has a metric. 
Let $H$ be an operator to act on $\HH$. 
Suppose that this space $\HH$ and this operator $H$ satisfy the following: 
\begin{itemize}
\item[(1)] 
There is a direct sum $\HH=\HH_- \oplus\HH_+$. 
\item[(2)]
There is an operator $Q:\HH\to\HH$ such that
$Q(\HH_+)\subset\HH_-$ and 
$Q(\HH_-)=0$. 
Let $Q^*$ denote the adjoint operator of $Q$. 
It holds that 
$Q^*(\HH_-)\subset\HH_+$ 
and 
$Q^*(\HH_+)=0$. 
Thus we have:

\hskip3cm
{\LARGE $0$}
$
\begin{array}{c}
                                  \\
\longleftarrow                    \\
{\text {\footnotesize $Q^*$}} \\
\end{array}
$
{\LARGE $\HH_+$}
$
\begin{array}{c}
{\text {\footnotesize $Q$}}        \\
\longrightarrow                    \\
\longleftarrow                     \\
{\text {\footnotesize $Q^*$}}
\end{array}
$
{\LARGE $\HH_-$}
$
\begin{array}{c}
{\text {\footnotesize $Q$}}       \\
\longrightarrow                   \\
                                  \\
\end{array}
$
{\LARGE $0$},
\vskip3mm

\f 
where $Q$ also denotes $Q|_{\HH_+}$ (resp. $Q|_{\HH_-}$) 
and $Q^*$ also denotes $Q^*|_{\HH_+}$ (resp. $Q^*|_{\HH_-}$). 

\item[(3)] 
 $H=QQ^*+Q^* Q$ 
 \end{itemize}

In literatures of Physics,  
$H$ is called the Hamiltonian, 
$\HH$ is called  a {\it sapce of states}. 
$\HH_+$ (resp. $\HH_-$) is called 
a space of bosonic (resp. fermionic ) states 
under some conditions (see the second Note from the following).  
$Q$ is a supercharge or 
an infinitesimal transformation of a supersymmetric transformation. 
Note that then the supersymmetric transformation is $e^{i\epsilon Q}$,  
where $\epsilon$ has some properties.@
See 
\cite{AlvarezGaume} 
\cite{WessBagger}  etc for detail.

Let $F$ be an operator to act on $\HH$ such that 
we have 
\nl  $F|q>=|q>$ and  $(-1)^F|q>=(-1)|q>$  
for any vector $|q>\in\HH_-$ and that 
\nl  $F|p>=0$ and $(-1)^F|p>=|p>$
for any vector $|p>\in\HH_+$, 
where note that $(-1)^F=e^{i\pi F}$. 

\noindent  
We define the {\it Witten index} 
for a set of $\HH$, $\HH_-$, $\HH_+$, $H$, $Q$ 
to be  dim$($Ker $Q|_{\HH_+})-$ dim$($Ker$Q^*|_{\HH_-}).$ 
The Witten index is often denoted by 
 tr$(-1)^{F}$ or tr$\{(-1)^{F}e^{-\beta H}\}$. 

\noindent 
{\bf Note.} This operation `tr' depends 
not only $\HH_+$, $\HH_-$, $F$ but also on $Q$. 
This `tr' is essentially same as `tr' in partition functions 
(see e.g. (4.5) (5.150) of \cite{Swanson}). 
\vskip3mm

Let $n_b=$dim$($Ker$Q|_{\HH_+})$ and $n_f=$dim$($Ker$Q^*|_{\HH_-})$. 
Then the Witten index is $n_b-n_f$. 
It is known that: 
Supersymmetry is broken if and only if $n_b=n_f=0.$ 
If supersymmetry is broken, then the Witten index is zero. 
That is, if the Witten index is nonzero, supersymmetry is not broken.

\noindent 
{\bf Note.} 
In order to make $\HH_+$ and $\HH_-$ more physics-like, we impose that 
the elements of $\HH_-$  (resp. $\HH_+$) 
satisfy (resp. do not satisfy ) the Pauli exclusion principle. 
(Some researchers only call the elements of $\HH_-$  (resp. $\HH_+$) 
fermions (resp. bosons) only under this condition. )
In the case  in \cite{AlvarezGaume}\cite{Witten} 
where the Witten index is the euler number, 
the product of differential forms gives such properties. 
However, 
in the case  in \cite{AlvarezGaume}\cite{Witten} 
where the Witten index is the signature, 
the product of differential forms does not give such properties. 
\vskip3mm


The papers \cite{AlvarezGaume} \cite{Witten} say that 
the Witten index is equal to 
the Atiyah-Singer index of the operator  
$Q:\HH_+\to\HH_-$ under some conditions. 
Recall that  the Atiyah-Singer index is 
dim$($Ker $Q|_{\HH_+})-$ dim$($Ker$Q^*|_{\HH_-}).$  
See \cite{AtiyahSinger}.

\f{\bf Note.}
\cite{AtiyahSinger} says that the Atiyah-Singer index theorem holds for 
elliptic operators of any order. 
\cite{AlvarezGaume} explanes that, 
for some kinds of oprators of the first order, 
the Atiyah-Singer index theorem is represented by a path integral. 
(The case of the first order elliptic operator is 
the most important case in today's physics.).  
\vskip3mm

The Witten index 
tr$\{(-1)^{F}e^{-\beta H}\}$ 
has path integral representation (see (2.5) (3.2) of \cite{AlvarezGaume})
$$\mathrm{tr}\{(-1)^{F}e^{-\beta H}\}=
\int_{\mathrm{PBC}}\DD\phi\DD\psi \mathrm{exp}[\int_o^\beta dt L].$$ 


\f
See e.g. \S5.5 of \cite{Swanson} and \cite{Alvarez} for 
APBC 
(anti-periodic boundary condition) 
and PBC.
 (periodic boundary condition). 
If $\HH_-$ (resp. $\HH_+$) does not have 
an anti-commutative (a commutative) product, 
the boundary condition is diffrent form the above. 
See e.g. (4.5) of \cite{AlvarezGaume}.

Partition functions tr$\{e^{-\beta H}\}$ are represented by path integrals. 
See (4.5) (4.12) (5.150) (5.151) (5.152) 
in  \cite{Swanson}. 
In the case of tr$\{(-1)^{F}e^{-\beta H}\}$,  
$(-1)^F$ makes the condition PBC.   
If we do not have $(-1)^F$, we have the condition APBC.

The domain of this path integral is 

\f 
$\SS=\{f|f:[0,t]\to E, f(0)=f(t)$ or $f(0)=-f(t)$ 
holds by the boundary condition. $\}$, 

\f where $t\in\R$, and 
$[0,t]=\{s|0\leqq s\leqq t,  t\in\R\}$, 
$E$ is the total space of a fiber bundle over a manifold $M$ 
with fiber $\G^n$. Here,  
$\G$ is the one dimensional vector space over complex Grassmann number 
and  $\G^n$ means $n$-dimensional vector space over complex Grassmann number. 
See \S5 of \cite{Swanson} for path integral representaions for fermions 
by using Grassmann variables, and ones for supersymmetry.

The notation $\int\DD x \DD y$ is derived from the fact that, 
in a chart $U\times \G^n$ of the $\G^n$ bundle, 
an element in $\SS$ is represented by 
$[0,1]\ni t\to(\phi, \psi)\in U\times\G^n$.

It holds that $\SS=\oplus_{i}A^i(M)$ for a $\G^n$-bundle. 
It holds that the Witten index is the euler number of 
$M$ for a combination of a Lagrangian and a $\G^n$-bundle.


\section{Our Witten indexes} \label{Our}

We make our Witten indexes. 

Let $K$ be an $n$-knot in $S^{n+2}$. 
Let $M^\mu$ be the $\mu$-fold branched covering space of $S^{n+2}$ along $K$
($\mu\in\N-\{1\}$).
See P.292 of \cite{Rolfsen} for branched covering spaces. 
We review it roughly in the second paragraph from the follwoing.

Let $X=S^{n+2}-{\rm{Int}}N(K)$. 
Let $\rho$ be the homomorphism $\pi_1 X\rightarrow H_1(X;\Z)$ 
which is obtained by the abelianization. 
Here, note that $H_1(X;\Z)\cong \Z$ 
and that 
the generator of $H_1(X;\Z)\cong \Z$ is determined by the orientation of 
$K$ and that of $S^{n+2}$. 
Let $\widetilde {X}^\infty$ be the covering space of 
$X$ associated with $\rho$.  
Let  $\alpha_\mu:\Z\to\Z_\mu$ be the natural epimorphism 
such that $\alpha_\mu$ carries the generator to the generator, 
where we fix the generator of $\Z_\mu$. 
Let $\rho_\mu=\alpha_\mu\circ\rho:\pi_1 X\to \Z_\mu$. 
Let $\widetilde {X}^\mu$ be the covering space of 
$X$ associated with $\rho_\mu$.

The branched covering space  $M^\mu$ has the following property: 
There is a continuous map $p:M^\mu\to S^{n+2}$ such that 
\begin{itemize}
\item[(1)] 
The map $p|_{p^{-1}(K)}: p^{-1}(K)\to K$ is a homeomorphism map. 

\item[(2)]
It holds that 
$[M^\mu-p^{-1}(K)]\cong {\rm{Int}}\widetilde {X}^\mu$ 
and 
$[S^{n+2}-K]\cong {\rm{Int}}X$. 

\item[(3)]
The map 
$p|_{[M^\mu-p^{-1}(K)]}:  
[M^\mu-p^{-1}(K)]\to [S^{n+2}-K]$ is 
same as 

\f 
the map 
$\rho_\mu|_{{\rm{Int}}\widetilde {X}^\mu}:  
{\rm{Int}}\widetilde {X}^\mu\to {\rm{Int}}X$. 

\end{itemize}

\f 
It is known that, if $\mu\neq\mu'$,  
then $M^\mu$ is not diffeomorphic to $M^{\mu'}$ in general.  
Note that the above branched covering space is 
a kind of generalization of 
branched covering spaces of Rieman surfaces in complex analysis.




\vskip3mm
We use the deRham cohomology (see \cite{deRham}).   
Let $M$ be an oriented closed manifold with a metric. 
Let $*$ denote the {\it Hodge (star) operator}  $A^i(M)\to A^{m-i}(M)$. 
We define an operator $d^*:A^i(M) \to A^{i-1}(M)$ so that 
$d^*=(-1)^i(*^{-1})d*$. Then it holds that $d^* d^*=0$. 
For $\alpha, \beta\in A^i(M)$, we define an inner product 
$< \alpha, \beta >=\int_M\alpha\wedge*\beta.$ 
Then it holds that $<d\alpha, \beta>=<\alpha, d^*\beta>.$
\vskip3mm

We define our Witten indexes for $n$-knots $K$.  
We define two kinds of indexes, 
\nl the $(2k, 2k+1;\mu)$-Witten index 
($0\leqq 2k \leqq 2k+1 \leqq n+2$, $k\in\Z$) and 
\nl the $(2k, 2k-1; \mu)$-Witten index 
($0\leqq2k-1\leqq2k\leqq n+2$, $k\in\Z$).

We firstly 
define the $(2k, 2k+1;\mu)$-Witten index ($0\leqq2k\leqq2k+1\leqq n+2$).
Take the $\mu$-fold branched covering space $M^\mu$ of $S^{n+2}$ along $K$. 
Give a metric on $M^\mu$. Let $A^i$ denote $A^i(M^\mu)$.

\vskip3mm
$(U^{2k}, U^{2k+1})= $ 
$$
\left\{ 
\begin{array}{ll}
\mbox{
$([\rm{Ker}(A^{2k-1}\stackrel{d^*}\gets A^{2k})]\otimes\C, 
[\rm{Ker}(A^{2k+1}\stackrel{d}\to A^{2k+2})]\otimes\C)$
}
& 
\mbox{if  $(2k, 2k+1)\neq(n+1, n+2), (0, 1)$}\\ 
\mbox{
$([\frac{A^0}{\rm{Ker}(A^{0}\stackrel{d}\to A^{1})}]\otimes\C, 
[\rm{Ker}(A^{1}\stackrel{d}\to A^{2})]\otimes\C)$
}
& 
\mbox{if $(2k, 2k+1)=(0, 1)$} \\ 
\mbox{
$([\rm{Ker}(A^{n}\stackrel{d^*}\gets A^{n+1})]\otimes\C, 
[\frac
{A^{n+2}}
{\rm{Ker}(A^{n+1}\stackrel{d^*}\gets A^{n+2})}]\otimes\C)$ 
}
& 
\mbox{if $(2k, 2k+1)=(n+1, n+2)$} 
\end{array}
\right. 
$$
\vskip3mm

\f 
Let
$\HH_+=U^{2k}$ and $\HH_-=U^{2k+1}$. 
We define 
$Q:\HH_+\to\HH_-$ 
by using $d$ naturally. 
We define $Q^*$ by using $d^*$ naturally.  
We define 
{\it the $(2k, 2k+1; \mu)$-Witten index}  
to be the Witten index for these.  

We next define the $(2k, 2k-1; \mu)$-Witten index ($0\leqq2k-1\leqq2k\leqq n+2$).

\vskip3mm
$(V^{2k}, V^{2k-1})= $ 
$$
\left\{
\begin{array} {ll} 
[\rm{Ker}(A^{2k+1}\stackrel{d}\gets A^{2k})]\otimes\C, 
[\rm{Ker}(A^{2k-1}\stackrel{d^*}\to A^{2k-2})]\otimes\C)
&
\mbox{
if $(2k,2k-1)\neq(n+2,n+1)$   
}\\
([\frac{A^{n+2}}
{\rm{Ker}(A^{n+2}\stackrel{d^*}\to A^{n+1})}]\otimes\C, 
[\rm{Ker}(A^{n+1}\stackrel{d^*}\to A^{n})]\otimes\C)
&
\mbox{
if $(2k,2k-1)=(n+2,n+1)$  } 
\end{array} 
\right.
$$
\vskip3mm

\f 
Let
$\HH_+=U^{2k}$, $\HH_-=U^{2k-1}$. 
We define $Q:\HH_+\to\HH_-$ by using $d^*$ naturally.  
We define $Q^*$ by using $d$ naturally. 
We define the 
{\it $(2k, 2k-1; \mu)$-Witten index} 
to be the Witten index for these.

We would explain a correspondence between 
our model and the real world. 
\begin{itemize}
\item[(1)] 
We give many pair of superpartners to a single $n$-knot . 
This situation corresponds to the fact that 
the real time-space is expected to have 
greater than one pair of superpartners. 
In our model, if we fix $\mu$, there are 
a pair of superpartners whose index is $(i,i+1)$ 
and 
another pair whose index is $(i,i-1)$. 
The letter $i$ appears in the two ways. 
This situation corresponds to the fact that:  
in the real time-space we expect the existence of 
a pair of superpartners 
whose index is $(\frac{1}{2},1)$ and 
another pair whose index is $(0, \frac{1}{2})$.  
If the particle has the index is $\frac{1}{2}$, 
there are two kinds of the index of the superpartner, that is 0 or 1.
(In the real time-space the indexes represent spin of particles.) 
\item[(2)] 
In order to make path integral representation for our Witten indexes, 
we restrict the space of states $A^i(M^\mu)$ 
and obtain a new space of states. 
Such restriction on  a space of states are often done in 
gauge theory in QFT and string theory 
(See e.g. Gupta-Bleuler treatment in \cite{GSW} etc.)

\item[(3)] 
A crucial point of the construction of our model 
is the fact that:  
a function which is an element of $\HH_+$ (resp.  $\HH_-$) 
is one from $M^{\mu}$ to $\C$, not from $S^{n+2}$, or not from $X$. 
We have the following situation in complex analysis 
and our construction is its generalization: 
An operation which we represent $z\to\sqrt{z}$ should be 
regarded as a function from a branched covering space 
of a Rieman surface 
not as a function from the base space of the branched covering space. 
\end{itemize}

\vskip3mm
From now we explain roughly that 
our Witten indexes satisfy the conditions in 
\S\ref{Introduction}.  

\f
By the definition,  our Witten indexes satisfy the following.

\begin{itemize}
\item[(1)] 
Our Witten indexes are topological invariants for $n$-knots. 

\item[(2)] 
The $(2k, 2k+1; \mu)$-Witten index is equal to 

$$
\left\{
\begin{array}{ll}
b_{2k}(M^\mu)-b_{2k+1}(M^\mu)
 &\mbox{if  $(2k, 2k+1)\neq(n+1, n+2), (0, 1)$} \\
b_{2k}(M^\mu)-b_{2k+1}(M^\mu)+1
 &\mbox{if $(2k, 2k+1)=(n+1, n+2)$} \\
b_{0}(M^\mu)-b_{1}(M^\mu)-1 
 &\mbox{if $(2k, 2k+1)=(0, 1)$} 
\end{array}
\right. 
$$

\item[(3)] 
The $(2k, 2k-1; \mu)$-Witten index is equal to 
$$
\left\{
\begin{array}{ll}
b_{2k}(M^\mu)-b_{2k-1}(M^\mu) 
 &\mbox{if $(2k,2k-1)\neq(n+2,n+1)$} \\
b_{2k}(M^\mu)-b_{2k-1}(M^\mu)-1 
 &\mbox{if $(2k,2k-1)=(n+2,n+1)$} 
\end{array}
\right. 
$$

\item[(4)] 
All of our Witten indexes are zero if and only if we have 
$H_l(M^\mu; \R)=H_l(S^{n+2}; \R)$ 
for all $\mu$ and all $l$.  
(Because (1) and (2) hold. )
\end{itemize}

For the trivial $n$-knots,  
$M^\mu$ is diffeomorphic to the $S^{n+2}$. 
Hence $H_l(M^\mu; \R)=H_l(S^{n+2}; \R)$ 
for all $l$ and $\mu$.  
Hence all of our Witten indexes are zero for the trivial $n$-knots.

\vskip3mm
We discuss relations between 
homology with twisted coefficients and our Witten index.
See e.g. Chap5 of \cite{DavisKirk} for homology with twisted coefficients 
for detail. 
Note that (co)homology with twisted coefficients is a kind of 
sheaf (co)homology.

\vskip3mm \noindent {\bf Definition. }
Let $Y$ be a CW complex. (Note that manifolds are CW coplexes.) 
Let $\widetilde Y$ be the universal covering space. 
Note that the chain complex 
 $C_i(\widetilde Y) (i\in\Bbb Z)$ 
 is regarded as a $\Bbb Z[\pi_1(X)]$-module by using 
 the covering transformation. 
 Given a  $\Bbb Z[\pi_1(X)]$-module $A$, 
 form the tensor product 
  $C_i(\widetilde Y)\otimes_{\Bbb Z[\pi_1(X)]} A$. 
  This is a chain complex whose differential is $\partial\otimes1$, 
where  $\partial$ is the differential of $C_i(\widetilde Y)$. 
The homology of this chain complex is called 
the {\it homology of $Y$ with local coefficients in $A$ }. 

\vskip3mm

Let $K$ be an $n$-knot in $S^{n+2}$. 
Take 
$X$, $\rho$, $\widetilde {X}^\infty$, $\rho_\mu$, and 
$\widetilde {X}^\mu$ as above. 
By using $\rho$ (resp. $\rho_\mu$), 
we can regard $\Bbb Z[\Bbb Z]$  (resp. $\Bbb Z[\Bbb Z_\mu]$) 
as a $\Bbb Z[\pi_1X]$ module.  
Let $H_i(X;\rho)$ (resp. $H_i(X;\rho_\mu)$ )
be the homology with the twisted coefficient such that 
Shapiro's lemma (see e.g. P.100 of\cite{DavisKirk}) says :  

\f 
$H_i(X; \rho)
\cong 
H_i(\widetilde{X}^\infty; \Z), \quad{\rm{and}}\quad 
 H_i(X; \rho)\otimes\R
\cong  
H_i(\widetilde{X}^\infty; \R) 
\quad{\rm{for}}\quad i\in\Z.  
$

\f 
$H_i(X; \rho_\mu)
\cong 
H_i(\widetilde{X}^\mu; \Z),   \quad{\rm{and}}\quad 
 H_i(X; \rho_\mu)\otimes\R
\cong  
H_i(\widetilde{X}^\mu; \R)    
\quad{\rm{for}} \quad i\in\Z  \quad\mu\in\N-\{1\}. 
$

By using the Meyer-Vietoris exact sequence, the Poincar\'e duality, 
and the universal coefficient theorem, 
the following three conditions are equivalent. Suppose $\mu$ is arbitrary.

\f (1) 
The $(i,j;\mu)$-Witten indexes ($i<k$ and $j<k$) are zero@
and 
the $(k-1,k;\mu)$-Witten index 
or 
the $(k,k-1;\mu)$-Witten index 
is nonzero.

\f (2) 
$H_i(M^\mu; \R)=0$ for $0< i< k$ 
and 
$H_k(M^\mu; \R)\neq0$.

\f (3) 
$H_1(X; \rho_\mu)\otimes\R=\R$, 
$H_i(X; \rho_\mu)\otimes\R=0$ for $1< i< k$ 
and 
$H_k(X; \rho_\mu)\otimes\R\neq0$.

\vskip3mm
We review the definition of the Alexander polynomial for $n$-knots. 
See 
\cite{LevineAlex} \cite{Rolfsen} 
for detail.
Let $K$ be an $n$-knot in $S^{n+2}$. 
Make $X$, $\widetilde {X}^\infty$, $\widetilde {X}^\mu$  as above.
Then we can regard 
$H_i(X; \rho)\cong H_i(\widetilde X^\infty; \Z)$ as $\Z[\Z]$-module 
by using the covering transformation 
associated with $\rho$.  
The $\Z[\Z]$-module 
$H_i(X; \rho)\cong H_i(\widetilde X^\infty; \Z)$ is 
called the {\it $i$-Alexander module} of an $n$-knot $K$ 
(over $\Z[\Z]$).  
Furthermore we can regard 
$H_i(X; \rho)\otimes\R\cong H_i(\widetilde X^\infty; \R)$ as $\R[\Z]$-module 
by using the covering transformation. 
The $\R[\Z]$-module 
$H_i(X; \rho)\otimes\R\cong H_i(\widetilde X^\infty; \R)$ is 
called the {\it $i$-Alexander module} of the $n$-knot $K$ 
(over $\R[\Z]$).  
Note that we can regard  
$H_i(X; \rho)\otimes\R\cong 
H_i(\widetilde X^\infty; \R)\cong 
(\R[\Z]/{\lambda_1})\oplus...\oplus(\R[\Z]/{\lambda_{\mu_i}}).$ 
We define the {\it $i$-Alexander polynomial} $\Lambda_{K,i}$ 
of the $n$-knot $K$@to be 
$\lambda_1\cdot...\cdot\lambda_{\mu_i}$.  
We must regard that $\Lambda_{K,i}(t)$ is same as 
  $r\cdot t^l\cdot \Lambda_{K,i}(t)$, 
  where $r\neq0$, $r\in\R$ and $l\in\Z$. 
If the $i$-Alexander polynomial $\Lambda_{K,i}(t)$ can be represented by  
$a_0+a_1\cdot t^1+...+a_{\nu_i}\cdot t^{\nu_i}$ ($\nu_i\in\N\cup\{0\}$, 
$a_0\neq0$, $a_{\nu_i}\neq0$.), 
then we say that 
the {\it degree} of the $i$-Alexander polynomial $\Lambda_{K,i}(t)$ 
is $\nu_i$. 
Note that the degree deg$\Lambda_{K,i}(t)$ 
of  the $i$-Alexander polynomial $\Lambda_{K,i}(t)$ coincides with 
the $i$ betti number of $\widetilde X^\infty$.  
(Recall that the $i$ betti number of $\widetilde X^\infty$ 
is the rank of the $\R$ vector space  $H_i(\widetilde X^\infty; \R)$.)

\vskip3mm
\f{\bf {Note}}: 
There are many other invariants for $n$-knots and $n$-links. 
See e.g. 
\cite{CochranOrr} \cite{Farber} \cite{Levinesimp} \cite{LevineOrr} 
\cite{O1}\cite{O2}\cite{O3}\cite{O4}\cite{O5}\cite{O6}\cite{O7}.

\vskip3mm
By using results on homology with twisted coefficients 
and those on covering spaces,  
we can find many results on our Witten indexes. 
Some of them are the following.

\begin{itemize}
\item[(1)] 
If one of our Witten indexes for an $n$-knot $K$ is nonzero, 
then one of  the betti numbers of $\widetilde X^\infty$ is nonzero. 
Hence one of the Alexander polynomials of $K$ is nontrivial. 
Hence one of the betti numbers of the homology $H_i(X;\rho)$ 
with the twisted coefficient defined by $\rho$ is nonzero.

\item[(2)] 
Let $K$ be a $1$-knot. Let $K$ be the trefoil knot. 
Then the (0,1;6)-Witten index is not zero. 
Let $K^{(n+1)}$ be an $(n+1)$-knot which is 
the spun-knot of an $n$-knot $K^{(n)}$($n\in\N$). 
Let  $K^{(1)}$ be the trefoil knot.  
(See  \cite{Rolfsen}, \cite{Zeeman} for spun-knot.) 
Then the (0,1;6)-Witten index for $K^{(n)} (n\in\N)$ is not zero.

Let $K$ be a simple $(4k+1)$-knot ($k>0$). 
(See e.g. 
\cite{Farber}
\cite{Levinesimp}
\cite{Rolfsen}
for simple knots.)
Suppose that one of Seifert matrixes for $K$ is 
$
\left(
\begin{array}{cc}
1&1\\
0&1
\end{array}
\right). 
$
Note that $K$ is not a trivial knot. 
Then the $(2k, 2k+1; 6)$-Witten index is not zero.

\item[(3)] 
Let $K$ be a $1$-knot. Let $K$ be the figure eight knot. 
Then all of our Witten indexes are zero. 
Let $K^{(n+1)}$ be an $(n+1)$-knot which is 
the spun-knot of an $n$-knot $K^{(n)}$($n\in\N$). 
Let  $K^{(1)}$ be the figure eight knot.  
Then all of our Witten indexes are zero.

Let $K$ be a simple $(4k+1)$-knot ($k>0$). 
Suppose that one of Seifert matrixes for $K$ is 
$
\left(
\begin{array}{cc}
1&1\\
0&-1
\end{array}
\right). 
$
Note that $K$ is not a trivial knot. 
Then all of our Witten indexes are zero. 

\end{itemize}

\vskip3mm\f{\bf {Note}}: 
In order to let the path integral make sense (in physics level), 
it might be better that  we suppose $\pi_1({M}^\mu)=1$ etc.
See e.g. \S7 of \cite{Alvarez}. 
Note that there are countably infinitely many nontrivial $n$-knots $K$ 
with $\pi_1({M}^\mu)=1$ for all $\mu$. 
For any simple $n$-knots $K$ and any $\mu$, 
it holds that $\pi_1({M}^\mu)=1$.

\vskip3mm\f{\bf {Note}}: 
 Our way gives an interpretation of  
the betti numbers of the homology with a twisted coefficient 
by using supersymmetric quantum theory 
not only  when we discuss $n$-dimensional knot 
but also when we discuss other manifolds.



\vskip3mm\f{\bf {Note}}: 
In our model we consider the branched covering $M^\mu$ 
along an $n$-knot $K$ over $S^{n+2}$. 
$M^\mu$ are many kinds of $(n+2)$-manifolds. 
The manifold $M^\mu$  may give a candidate for a fiber 
if we regard our time-space as 
the fiber bundle $\R^4\x M$ with the base sapce $\R^4$ and with the fiber 
$M$. 
In this model we suppose that 
we can usually observe $\R^4\x M$ as $\R^4$ since $M$ is very small.   
Anyway we have the following if $M^\mu$ is a 6-dimensional or 7-dimensional 
manifold.

\begin{itemize}

\item[(4)] 
Let $K$ be a 4-dimensional knot in $S^6$. 
Then all of our supersymmetries 
are broken if the all Alexander polynomials of $K$ are zero.

\item[(5)] 
Let $K$ be a 5-dimensional knot in $S^7$. 
Then it holds: 
the $(4,3;q)$-Witten index for any $q$ is zero and 
there are countably infinitely many 5-knots such that 
supersymmetry assciated with the $(4,3;q)$-Witten index are 
broken (resp. not broken). 

\end{itemize}

\noindent{\bf Acknowledgment.}
The author would like to thank 
Prof. Tohru Eguchi, Prof. Akio Sugamoto, Prof. Tomoyoshi Yoshida, 
Prof. Yutaka Matsuo, Dr. Kentaro. Hori, Dr. Yukiko Konishi 
for their encouragement and for their interest. 

\end{document}